\documentclass{INTERSPEECH2023}
\usepackage{comment}
\usepackage{amsmath,graphicx,multirow,subcaption}
\usepackage{color, colortbl}
\usepackage[usenames,dvipsnames]{xcolor}
\usepackage{makecell}

\newcommand{\COMMENT}[1]{}

\newcommand{\df}[1]{{\bf [\textcolor{red}{Daniele. #1}{\bf ]}}}

\interspeechcameraready

\title{AN INVESTIGATION OF THE COMBINATION OF REHEARSAL AND KNOWLEDGE DISTILLATION IN CONTINUAL LEARNING FOR SPOKEN LANGUAGE UNDERSTANDING}
\name{Umberto Cappellazzo$^1$, Daniele Falavigna$^2$, Alessio Brutti$^2$}
\address{
  $^1$University of Trento, Trento, Italy\\
  $^2$Fondazione Bruno Kessler, Trento, Italy}
\email{umberto.cappellazzo@unitn.it, \{falavi,brutti\}@fbk.eu}

\begin{document}

\maketitle
 
\begin{abstract}
Continual learning refers to a dynamical framework in which a model receives a stream of non-stationary data over time and must adapt to new data while preserving previously acquired knowledge. Unluckily, neural networks fail to meet these two desiderata, incurring the so-called catastrophic forgetting phenomenon. Whereas a vast array of strategies have been proposed to attenuate forgetting in the computer vision domain, for speech-related tasks, on the other hand, there is a dearth of works. In this paper, we consider the joint use of rehearsal and knowledge distillation (KD) approaches for spoken language understanding under a class-incremental learning scenario. We report on multiple KD combinations at different levels in the network, showing that combining feature-level and predictions-level KDs leads to the best results. Finally, we provide an ablation study on the effect of the size of the rehearsal memory that corroborates the efficacy of our approach for low-resource devices.

\end{abstract}

\section{Introduction}
\label{sec:intro}

Spoken Language Understanding (SLU) plays a crucial role in countless speech-related applications, such as virtual assistants and home devices \cite{tur2011spoken}. Its main purpose is to extract relevant information from a spoken utterance. Intent classification is a core problem of every SLU system and involves the identification of the correct intent associated with a specific utterance. Although several works have targeted this problem with exceptional results \cite{lugosch2019speech,seo2022integration}, it still misses an in-depth study under a continual learning (CL) setting, in which the entire data is not available to the system at once, but it is spread over a sequence of tasks.

CL is a machine learning field that studies how to mitigate catastrophic forgetting, defined as the proclivity of neural networks to fit the new data distribution at the expense of the knowledge that has been already learned \cite{mccloskey1989}. This setting is in line with practical applications that require a model to be robust to unforeseen drifts in the input data distribution. 
CL approaches proposed in the literature can be divided into three categories: {\em a)} methods based on a regularisation loss \cite{li2017learning, kirkpatrick2017overcoming} that prevents the parameters from changing widely, {\em b)} rehearsal methods based on the replay of historical data \cite{hayes2021replay, rebuffi2017icarl} (either training samples \cite{lopez2017gradient} or model weights \cite{rosenfeld2018incremental}), and {\em c)} 
methods that modify the architecture of the model \cite{yan2021dynamically}.

In this work, we consider the problem of intent classification applied to a class-incremental learning (CIL) scenario, whereby the intents are distributed into several tasks, and the model has to correctly learn them sequentially. The main hindrance of such a scenario is twofold: 1) the model has access to only the intents from the current task, and 2) the task identifier is unknown to the model, yet it must be predicted along with the intents.

Many works \cite{lopez2017gradient,rebuffi2017icarl,buzzega2020dark} have proven that an extremely  effective approach to reducing catastrophic forgetting relies on the usage of a set of rehearsal data, chosen at the end of each task and stored in a rehearsal memory, that contribute to the loss function together with those in the actual task. The additional use of a distillation loss \cite{li2017learning}, which alone falls through in a CIL scenario, is not always beneficial to the model, as pointed out in \cite{belouadah2019il2m,masana2022class}, who contend that it can even lead to a deterioration in the performance. Also, their mutual interaction is scarcely investigated for speech as well as for other modalities (e.g., vision).

For this reason, we investigate the interrelationship between applying the knowledge distillation (KD) \cite{gou2021knowledge} at different levels in the model, namely in the predictions and in the feature space, and the rehearsal approach. We demonstrate that the joint use of predictions-level and feature-level KDs leads to the best results.

Our contributions to the CL problem are the following: i) we define a CIL scenario for SLU over the Fluent Speech Command dataset \cite{lugosch2019speech}; ii) we provide a thorough analysis of the combination of rehearsal and KDs for 4 CL strategies, and we prove its efficacy in our scenario; iii) we point out that a careful design of the KD weights is crucial for obtaining optimal results, and we foster the CL community to place more emphasis on this aspect; iv) we provide an ablation study about the size of the rehearsal buffer, and we conclude that our approach attains larger gains for smaller sizes, thus making it appealing for low-resource devices.

\section{Related work}
\label{sec:related work}

CL strategies can be categorized into three main groups \cite{de2021continual, parisi2019continual}: regularization, rehearsal, and architectural approaches.

\textit{Regularization} approaches mitigate catastrophic forgetting by supplying the standard cross-entropy (CE) loss with regularization terms to avoid abrupt changes in the model weights. Learning without forgetting (LwF) \cite{li2017learning} employs a weighted knowledge distillation loss that forces the outputs of the model to be similar to those obtained by the model trained in the previous task. The work by \cite{kirkpatrick2017overcoming} resorts to the Fisher information matrix to estimate the importance of the model weights and protect them afterward to prevent forgetting, while \cite{douillard2020podnet} advance a spatial-based distillation loss computed for every intermediate layer of the model.

\textit{Rehearsal} or \textit{Experience Replay} methods keep in memory some of the past data  
to mitigate forgetting. A key aspect lies in the selection strategy for retaining past data. The simplest, but relatively effective, approach  randomly chooses some samples from the last task \cite{chaudhry2018riemannian}. ICaRL \cite{rebuffi2017icarl} fosters the samples whose features are the closest to their moving center of gravity. Gradient Episodic Memory (GEM  \cite{lopez2017gradient}) attenuates forgetting by projecting the gradient update along a direction that does not interfere with the previously acquired knowledge. \cite{gal2017deep} select the rehearsal samples by maximizing the mutual information between the predictions and the posterior of the model's parameters using Monte Carlo dropout.

Finally, \textit{Architectural} methods apply modifications to the network architecture, such as adding layers or freezing specific parts of the model, to handle new incremental tasks. An example is \cite{yan2021dynamically}, where, at each new task, a novel feature extractor is instantiated, while the previous one is frozen, and pruning is applied to shrink the model. An alternative that has been in vogue recently relies on prompt learning, namely a small portion of new parameters are appended to all data of a new task and they are learned, whilst the network is kept frozen. \cite{liu2022incremental} apply this paradigm for 
class-incremental event detection.
These methods, although successful, do not scale to the number of seen tasks.

Although the literature is mainly related to computer vision, CL has also been investigated in the speech domain. For example, \cite{huang2022progressive} address a KWS incremental-learning task, creating a sub-network for each new task and keeping in memory the processed features from the past tasks. It is also worth noting the use of CL in Automatic Speech Recognition (ASR). The work in \cite{yang2022online} proposes an online GEM method for model updates together with a selective sampling strategy. 

For SLU, we find few works that only consider a domain-incremental CL scenario: \cite{shen2019progressive} propose a progressive architecture for the slot-filling task that expands the network for each new task; \cite{mi2020continual} consider the combination of rehearsal and regularization techniques for natural language generation. Nevertheless, to the best of our knowledge, we are the first to explore SLU in a CIL scenario, in particular, we study the adoption of the KD at features and predictions levels, and applied to only the rehearsal data or the rehearsal data plus the current data.

\section{Proposed approach}
\label{sec:methodology}

\vspace{-0.5cm}
\begin{figure}[htb]
\centering
\includegraphics[width=8cm]{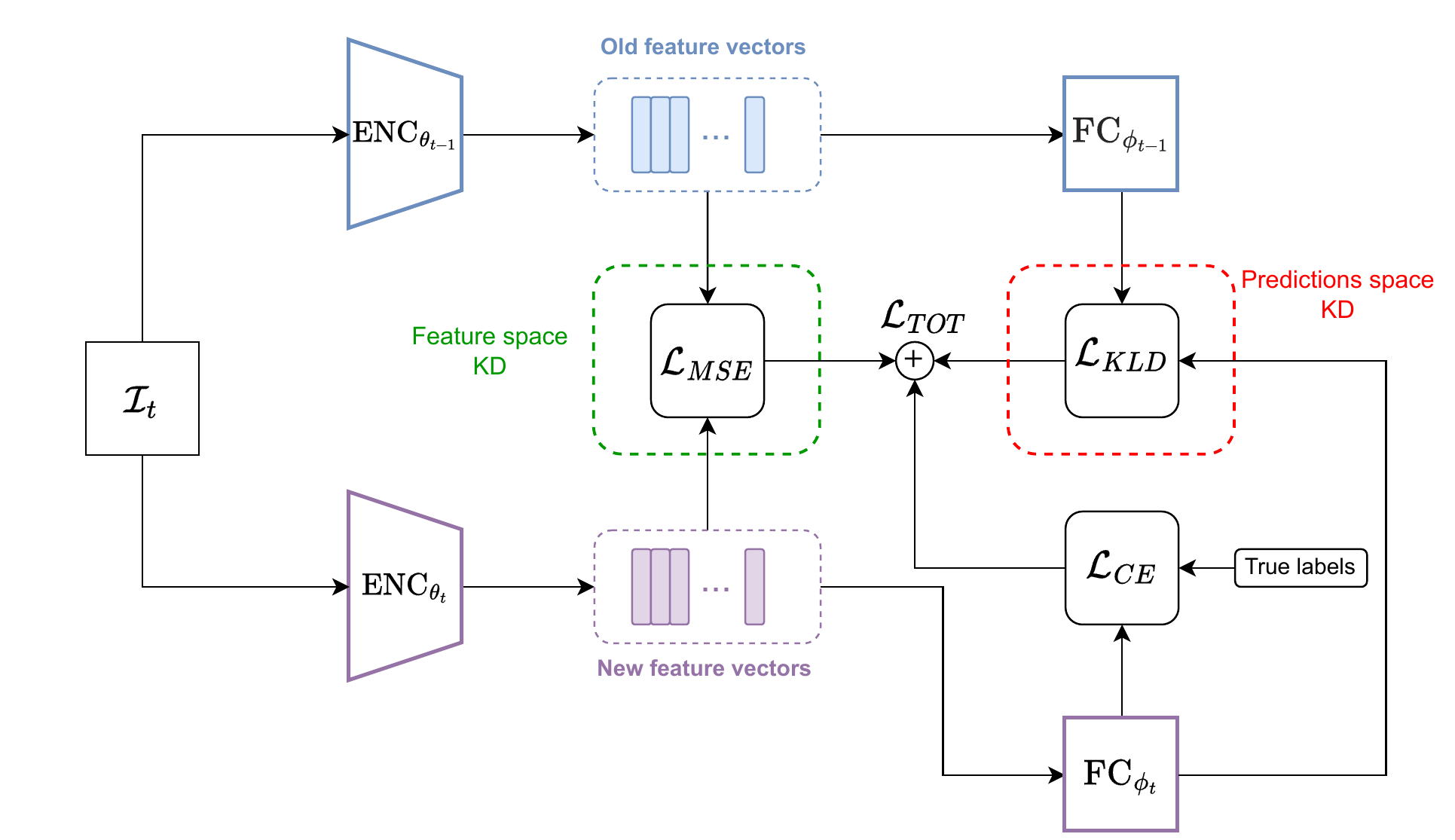}
\caption{Complete overview of our proposed approach.}
\label{fig:model_pipeline}
\end{figure}

In a CL setting, a classification model, which comprises a multilayered feature extractor $\text{ENC}_{\theta}$ and a classifier $\text{FC}_\phi$ (parameterized by $\theta$ and $\phi$, respectively), is trained over a sequence of $T$ distinct training phases, that is $\mathcal{D}=\{\mathcal{D}_0,\ldots,\mathcal{D}_{T-1}$\}. The dataset $\mathcal{D}_t$ related  to the $t^{th}$ training phase is interpreted  as a task defined by audio signals $\mathcal{X}_t$
and associated class labels $\mathcal{Y}_t$, i.e. $\mathcal{D}_t=(\mathcal{X}_t,\mathcal{Y}_t)$. In CIL scenarios all task label sets are mutually exclusive, i.e. $\mathcal{Y}_i\cap\mathcal{Y}_j=\O, i\neq j$.

At the end of task $t-1$ we select a set of data $\mathcal{R}_{t-1}\subset\mathcal{D}_{t-1}$ for the rehearsal memory. Then, all the rehearsal data, from task $0$ to task $t-1$, $\mathcal{R}_0^{t-1}= \{\mathcal{R}_0, \dots, \mathcal{R}_{t-1}\}$ are joined with the training data $\mathcal{D}_{t}$ in order to train the model  for the $t^{th}$ task. A naive CL strategy optimizes the CE loss computed over $\mathcal{D}_t\cup\mathcal{R}_0^{t-1}$:

\begin{equation}
\hspace{-1cm}\mathcal{L}^t_{CE}=-\sum_{(x,y)\in\mathcal{D}_t\cup\mathcal{R}_0^{t-1}}\log(p[y|x;(\theta_{t},\phi_{t})]),
\label{eq:ce}
\end{equation}
where $p[y|x;(\theta_{t},\phi_{t})]$ is the output probability distribution of the model given the parameters $\theta_t$ and $\phi_t$ at task $t$. 

A common approach is to further regularize the model adaptation through a KD loss. In this paper, we experiment with two different distillation terms in combination with the CE loss. The first one is the Kullback Leibler Divergence (KLD) between the output probability distribution at task $t$ and the distribution predicted with the model trained at task $t-1$, i.e.:
\begin{equation}
\mathcal{L}_{KLD}^t = \sum_{(x,y)\in\mathcal{I}_t}p[y|x;(\theta_{t-1},\phi_{t-1})]\log(p[y|x;(\theta_{t},\phi_{t})]).   
\label{eq:kld}
\end{equation}
In the equation above $\mathcal{I}_t$ represents the training set for task $t$, consisting of only the rehearsal data ($\mathcal{I}_t=\mathcal{R}_0^{t-1}$), or the union of the rehearsal and current data of task $t$ ($\mathcal{I}_t=\mathcal{D}_t\cup\mathcal{R}_0^{t-1}$). 
The second regularization term is given by the mean squared error (MSE) loss between the output of the model encoder at tasks $t-1$ and $t$, i.e.:
\begin{equation}
\mathcal{L}_{MSE}^t=\sum_{x\in\mathcal{I}_t}\Vert \text{ENC}_{\theta{_{t-1}}}(x)-\text{ENC}_{\theta_{t}}(x)\Vert^2.
\label{eq:mse}
\end{equation}
Also in this case we experiment with both $\mathcal{I}_t=\mathcal{R}_0^{t-1}$ and $\mathcal{I}_t=\mathcal{D}_t\cup\mathcal{R}_0^{t-1}$. 
The total loss to optimize in each task $t$ is therefore a linear combination of the CE loss in eq.~\ref{eq:ce} and the regularization losses in eqs.~\ref{eq:kld} and ~\ref{eq:mse}: 
\begin{equation}
\mathcal{L}^t_{TOT}=\lambda_{CE}\mathcal{L}^t_{CE}+\lambda_{KD}\mathcal{L}^t_{MSE} +  \lambda_{KD}\mathcal{L}^t_{KLD}.
\label{eq:ltot}
\end{equation}
Figure~\ref{fig:model_pipeline} shows a schematic illustration of the proposed CL approach.

\section{Experiments}
\label{sec:experiments}
\subsection{CIL definition for FSC and rehearsal memory}

We evaluate our proposed approach on the Fluent Speech Commands (FSC) dataset \cite{lugosch2019speech}. 
FSC includes 30,043 English utterances, recorded at 16 kHz. The dataset provides 248 different utterances mapped in 31 different intents. There is only one intent per utterance. To give an example, the intent \textit{increase\_heat\_kitchen} is associated with the utterance ``\textit{turn up the temperature in the kitchen}''. We split the dataset into train, validation, and test sets with a ratio of 80:10:10 as proposed in \cite{lugosch2019speech}.

To define the CIL scenario, we partition the FSC dataset into 10 disjoint tasks, where the first task comprises 4 unique intents and the subsequent 9 tasks contain 3 intents. The order of the intents is random, and we have not observed significant change by considering different random orders. 

Concerning the rehearsal memory, its entire capacity is not exploited since the beginning, but each class has a pre-allocated space that is used when that class is seen for the first time. In this way, we avoid a possible imbalance among the classes between the first and the last tasks.  
\vspace{-0.3cm}
\subsection{Model architecture}
\label{sec:implementation}

\begin{figure}[htb]

\centering
\begin{subfigure}[b]{\columnwidth}
         \centering
         \includegraphics[width=\textwidth]{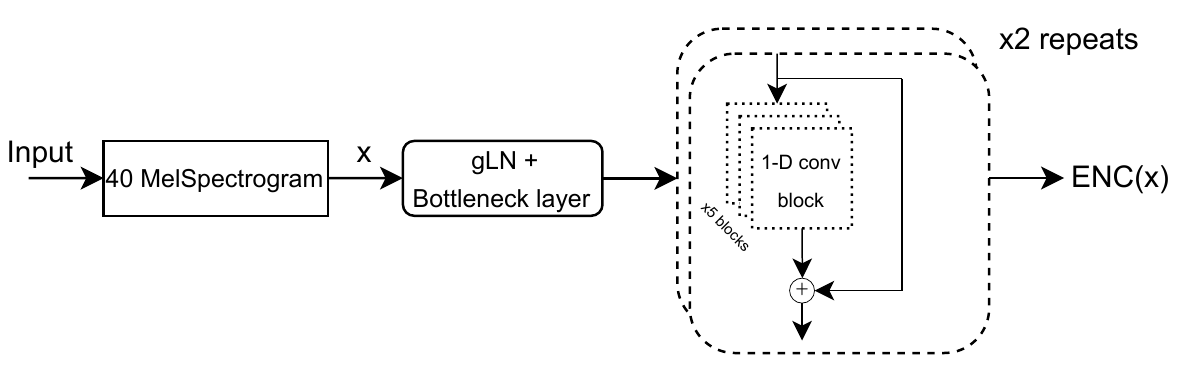}
         \caption{The TCN architecture.}
         \label{fig:sequential}
     \end{subfigure}
    \begin{subfigure}[b]{\columnwidth}
         \centering
         \includegraphics[width=\textwidth]{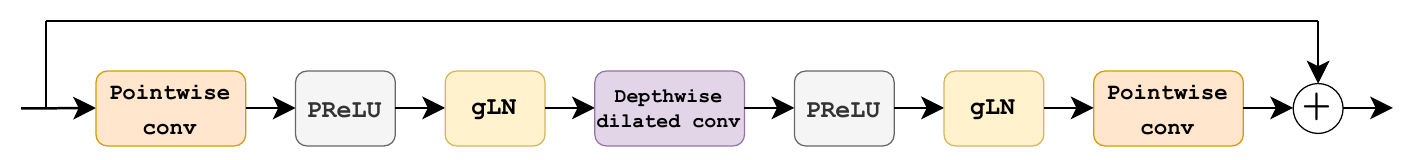}
         \caption{The Conv1D residual block.}
         \label{fig:conv1d}
     \end{subfigure}
\caption{The block diagram of the TCN encoder performing feature extraction and the corresponding Conv1D block structure.}
\label{fig:nn-arch}
\end{figure}

The deep feature extractor we employ in the experiments is depicted in Figure~\ref{fig:nn-arch}. It is inspired by the temporal convolutional network (TCN) used in the separation block of Conv-Tas-Net \cite{luo2019conv}, a recently proposed model for speech separation. We observe that the KD strategies we propose are \textit{architecture-agnostic}, so they do not rely on the underlying architecture. It would be possible to substitute the TCN with any other deep architecture (e.g., transformer-based encoder).

\begin{table}[!htb]
\centering 
\caption{List of the hyperparameters of the TCN.}
\begin{tabular}{l|c}
\Xhline{2\arrayrulewidth}
\textbf{Hyperparameter} & \textbf{Value} \\
\hline
Input channels & 40 \\
Hidden channels & 128 \\
Output channels & 64 \\
\# 1-D conv blocks & 5 \\
\# Repetitions & 2 \\
Kernel size for the depthwise conv & 3 \\
\Xhline{2\arrayrulewidth}
\end{tabular}
\label{table:tcn_hyperparams}
\end{table}

The network takes as input 40 Mel-spaced log filter-banks, computed using a sliding window of length 25 ms, with 10 ms stride. Then, it applies a global layer normalization (gLN) and a bottleneck layer (1x1 conv block) that maps the input features into 64 channels. The input layer is followed by 2 repetitions of 5 consecutive 1-D dilated convolutional residual blocks. Each residual block is formed by two symmetrical pipelines surrounding a depth-wise separable convolutional layer that maps the 64 bottleneck features into 128 channels. A residual branch connects the original input to the output. Mean pooling is applied to the output of the last block, followed by gLN and a linear layer. A softmax activation layer gives the final class scores. 

We train the TCN model for 50 epochs per task with Adam optimizer, with a learning rate equal to $5e^{-4}$. The CIL scenario is implemented with the Continuum library \cite{douillard2021continuum}, and the rest of the code is based on PyTorch. For each experiment, we use one Quadro RTX A5000. One experiment with memory size = 930 and with the iCaRL method requires around 1.15 hours, whereas the GEM method around 6 hours. Table~\ref{table:tcn_hyperparams} reports the whole set of hyperparameters. The code is available online\footnote{https://github.com/umbertocappellazzo/CL\_SLU}.
\subsection{Distillation weights}
The selection of the KD weights deserves special attention. A common choice for the $\lambda_{KD}$ weight is $\frac{n}{n+m}$, where $n$ is the number of old (seen) classes and $m$ is the number of new classes \cite{wu2019large,zhao2020maintaining}. This choice was originally proposed for works that used only KD as CL strategy, and it gives more and more importance to $\lambda_{KD}$ over time because the past model retains the knowledge from more and more past classes. The subsequent works that considered KD and rehearsal together, adhered to this choice. Nonetheless, we speculate that relying on this option gives worse results. 
When we use both KD and rehearsal approaches applied to rehearsal and current data ($\mathcal{I}_t=\mathcal{D}_t\cup\mathcal{R}_0^{t-1}$), the importance of the past model is damped by the fact that the current model sees the rehearsal data, so we still would like $\lambda_{KD}$ to increase, but at a slower pace, and this can be accomplished by using a log function. When we use the KD applied only to the rehearsal data ($\mathcal{I}_t=\mathcal{R}_0^{t-1}$), we give it a weight proportional to the fraction of rehearsal data in the mini-batch. Since this number is too small during the first tasks, we apply the square root operation to enlarge it. 
Ultimately, we set $\lambda_{KD}$ as follows:

\begin{equation}
\lambda_{KD} =  
\begin{cases}
\log(1+\frac{n}{n+m}) & \text{if }  \mathcal{I}_t=\mathcal{D}_t\cup\mathcal{R}_0^{t-1} \\
\sqrt{\frac{b_{rehe}}{b_{all}}} & \text{if }  \mathcal{I}_t=\mathcal{R}_0^{t-1}
\end{cases}
\label{eq:KD}
\end{equation}
where $b_{rehe}$ counts the number of rehearsal data in the current mini-batch, and $b_{all}$ is the current mini-batch size. We found empirically that using $\lambda_{KD}$ as defined in eq. \ref{eq:KD} brought about a 1\% to 2\% improvement in the accuracy. This study suggests that a careful choice of the KD weights is essential. 

\begin{table*}[ht]
    \centering
    \caption{Intent classification accuracy with 930 samples in the rehearsal memory, using different distillation strategies. The highest accuracies are reported in bold while we use italics for the best KD of each CL method.}
    \begin{tabular}{llcccccccc}
    \hline
    \multicolumn{2}{l}{\textbf{Baselines}} & \multicolumn{2}{c}{\textbf{last acc}} &
    \multicolumn{2}{c}{\textbf{avg acc}} \\
    \hline
    \multicolumn{2}{l}{Offline} & \multicolumn{2}{c}{0.985} &
    \multicolumn{2}{c}{-} \\
    \multicolumn{2}{l}{Finetuning} & \multicolumn{2}{c}{0.073} &
    \multicolumn{2}{c}{0.267} \\
    \multicolumn{2}{l}{Pred. KD (no rehe)} & \multicolumn{2}{c}{0.080} &
    \multicolumn{2}{c}{0.272} \\
    
    \Xhline{2\arrayrulewidth}
    \textbf{Feat. KD} & \textbf{Pred. KD}  & \multicolumn{2}{c}{Random} & \multicolumn{2}{c}{Closest\_to\_mean}  &\multicolumn{2}{c}{iCaRL \cite{rebuffi2017icarl}} &
    \multicolumn{2}{c}{GEM \cite{lopez2017gradient}}\\
    \textbf{data} &\textbf{data} & \textbf{last acc} & \textbf{avg acc}& \textbf{last acc} & \textbf{avg acc}&  \textbf{last acc} & \textbf{avg acc} & \textbf{last acc} & \textbf{avg acc}\\
    \hline
    - & - & 0.660 & 0.720 & 0.650& 0.694 & 0.682& 0.740 & 0.573 & 0.710 \\
    \hline
   \rowcolor{green}
    \multicolumn{10}{c}{\textbf{Feature space KD}}\\
    \hline
   $\mathcal{R}$ & - & 0.737 & 0.779 &  0.728& 0.740 & 0.789 & 0.802 & \it{0.773} & 0.789  \\
   $\mathcal{D}\cup\mathcal{R}$  & - & 0.594 &0.643 & 0.562 &0.609 &  0.600 &0.643  & 0.710 & 0.714  \\
    \hline
    \rowcolor{pink}
    \multicolumn{10}{c}{\textbf{Predictions space KD}}\\
    \hline
    - & $\mathcal{R}$ & 0.676 & 0.736 & 0.632 &0.690 & 0.662 & 0.726 & 0.624 & 0.735 \\
    - &  $\mathcal{D}\cup\mathcal{R}$ & 0.757& 0.764 & 0.690& 0.717 & 0.780& 0.795 & 0.600 & 0.710 \\
    \hline
    \rowcolor{lightgray}
    \multicolumn{10}{c}{\textbf{Double KDs}}\\
    \hline
    $\mathcal{R}$ & $\mathcal{R}$ & 0.752& 0.770& 0.728 & 0.739 & 0.788 & 0.787 & 0.764 & \it{0.799}  \\
    $\mathcal{R}$ & $\mathcal{D}\cup\mathcal{R}$ & {\it 0.771} & {\it 0.796} & {\it 0.729}& {\it 0.740}  & {\bf 0.811} & {\bf 0.812} & 0.751 & 0.796 \\
    \hline
    \end{tabular}
    \label{tab:tabel_930_new}
\end{table*}

Eq. \ref{eq:ltot} changes depending on the considered experiment. When we do not apply any KD loss, the weights boil down to $\lambda_{KD} = 0$, $\lambda_{CE} = 1$ (in practice, only the CE loss is used). When we use the KD in the feature space only, the KLD loss is not present, $\lambda_{KD}$ follows eq. \ref{eq:KD}, and $\lambda_{CE}$ = 1 - $\lambda_{KD}$. If we use the KD in the predictions space, the same as before applies with the KLD loss and the MSE loss inverted. Lastly, when both the KLD loss and the MSE loss are employed, their coefficient $\lambda_{KD}$ follows eq. \ref{eq:KD}, and $\lambda_{CE}$ is set to 1.
\subsection{Results}
\label{sec:results}

Table~\ref{tab:tabel_930_new} reports the intent classification accuracy for different KD strategies in combination with 4 \COMMENT{state-of-the-art} CL approaches, i.e. a rehearsal approach with 3 different sample selection strategies (random, iCaRL \cite{rebuffi2017icarl}, and ``closest\_to\_mean'', where the samples which are closest to their class mean in the feature space are chosen), and GEM \cite{lopez2017gradient}. The rehearsal memory size is 930 (around 4\% of the dataset size). We consider 2 random class orders, and for each, we run 4 experiments and take the average. We use 2 metrics to test the efficacy of each strategy: the average accuracy (\textbf{avg acc}), which is defined as the average of the accuracies after each task, and the accuracy after the last task (\textbf{last acc}). For better stability, the accuracy of each task is the average of the last 5 epochs.

In the upper part of Table~\ref{tab:tabel_930_new} we include the results for: {\em i)} the offline upper bound (i.e., no incremental learning), which is in line with the current state-of-the-art methods on the FSC dataset; {\em ii)} the results obtained with the naive fine-tuning method, and {\em iii)} the results we achieve applying solely the KD in the predictions space without rehearsal.

The lower part of the table shows the accuracies when rehearsal data are employed. The rows show the accuracies for the cases in which the distillation is performed at the feature level, predictions level, and both levels, respectively. For each configuration, the table also reports the performance when distillation is applied to either rehearsal data alone (denoted with $\mathcal{R}$ in the table) or to the union  of rehearsal and actual task data (denoted with $\mathcal{D}\cup\mathcal{R}$).

When we endow the model with the KD in the feature space, the reliance on only the rehearsal data improves both the average accuracy and the last accuracy. On the contrary, the joint use of $\mathcal{D}\cup\mathcal{R}$ deteriorates the performance. This can be explained by observing that if we use $\mathcal{D}\cup\mathcal{R}$, we are forcing the current model, $\theta_t$, to produce feature representations that are similar to the ones obtained with the previous model, $\theta_{t-1}$. Whereas this is desirable for the rehearsal data (the previous model has been trained on them), this is not the case for the new data, since we want our model to learn in the actual task new clusters which should be far apart from the past clusters.

Considering the KD in the predictions space, instead, we witness a trend inversion. The use of $\mathcal{D}\cup\mathcal{R}$ achieves better results than just using the data in the memory, albeit the difference is not as pronounced as for the feature-space KD. This can be explained by the fact that since in the predictions space we deal with probability distributions, the new and teacher models produce values for both new and past classes, thus it is more convenient to apply the KD to $\mathcal{D}\cup\mathcal{R}$ (e.g., $\mathcal{D}$ and $\mathcal{R}$ act as negative samples for the past and new classifiers, respectively). It is also worth noting that in almost all cases the feature-level KD attains slightly better results than its predictions counterpart. We point out that GEM achieves slightly better results when only rehearsal data are considered, and this may be because it already employs a regularization on the gradients using only the rehearsal data.
\vspace{-0.3cm}
\begin{figure}[htb]
\centering
\includegraphics[width=8cm]{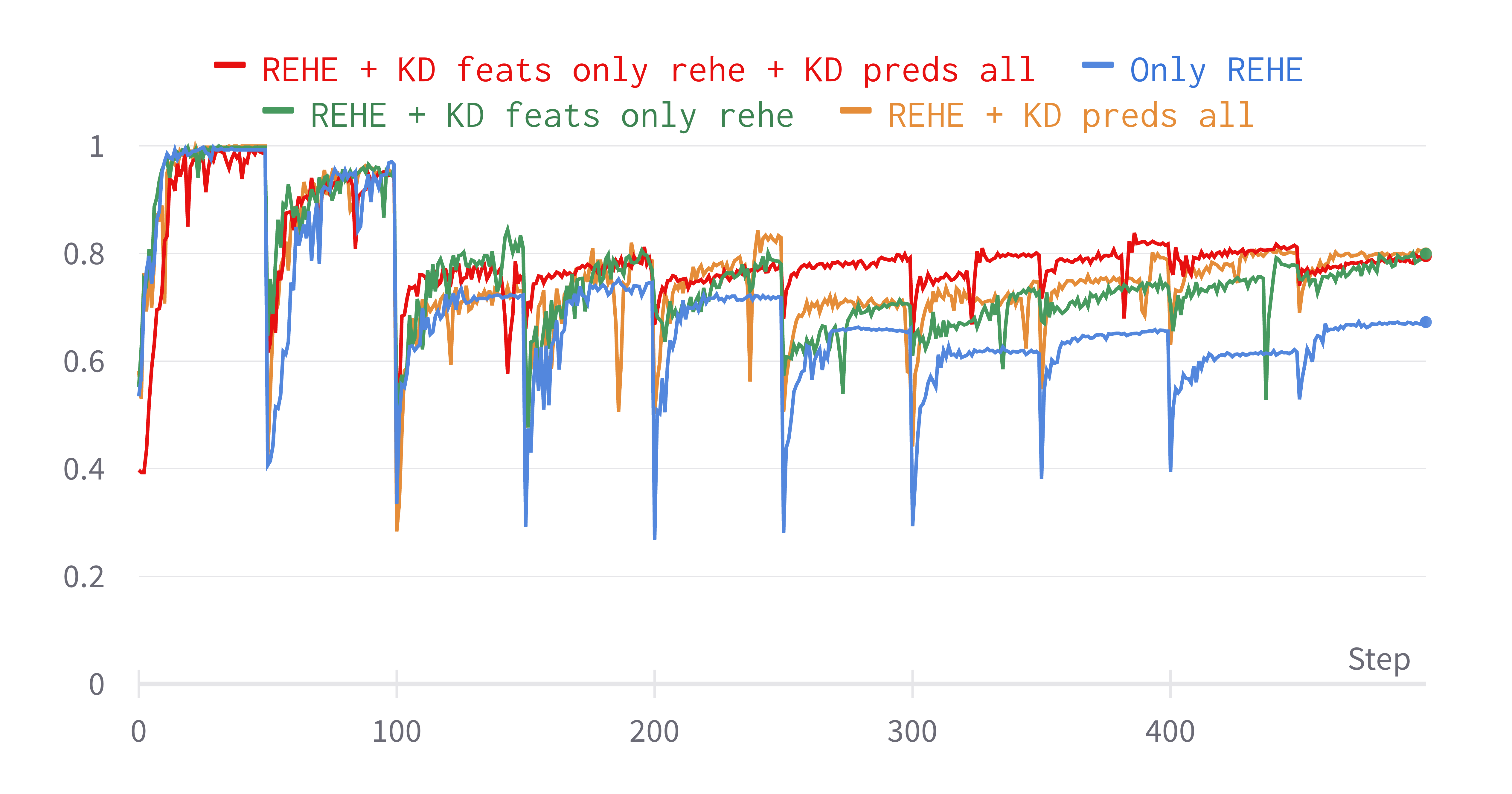}
\caption{Trend of the avg accuracy for 4 different combinations of the iCaRL method. 
Each task has 50 steps (i.e., epochs).}
\label{fig:avg_acc_trend}
\end{figure}

The last two rows of Table \ref{tab:tabel_930_new} consider the combination of feature-level and predictions-level KDs (the configurations with $\mathcal{D}\cup\mathcal{R}$ in the feature space are not considered since we have shown they are harmful to the model). The use of the features-space KD applied to $\mathcal{R}$ in conjunction with the predictions-space KD applied to $\mathcal{D}\cup\mathcal{R}$ gives the best results (0.811 and 0.812 for the last acc and avg acc by iCaRL, respectively), proving the effectiveness of integrating both KDs.
\begin{figure}[h]
\centering
\includegraphics[width=7.5cm]{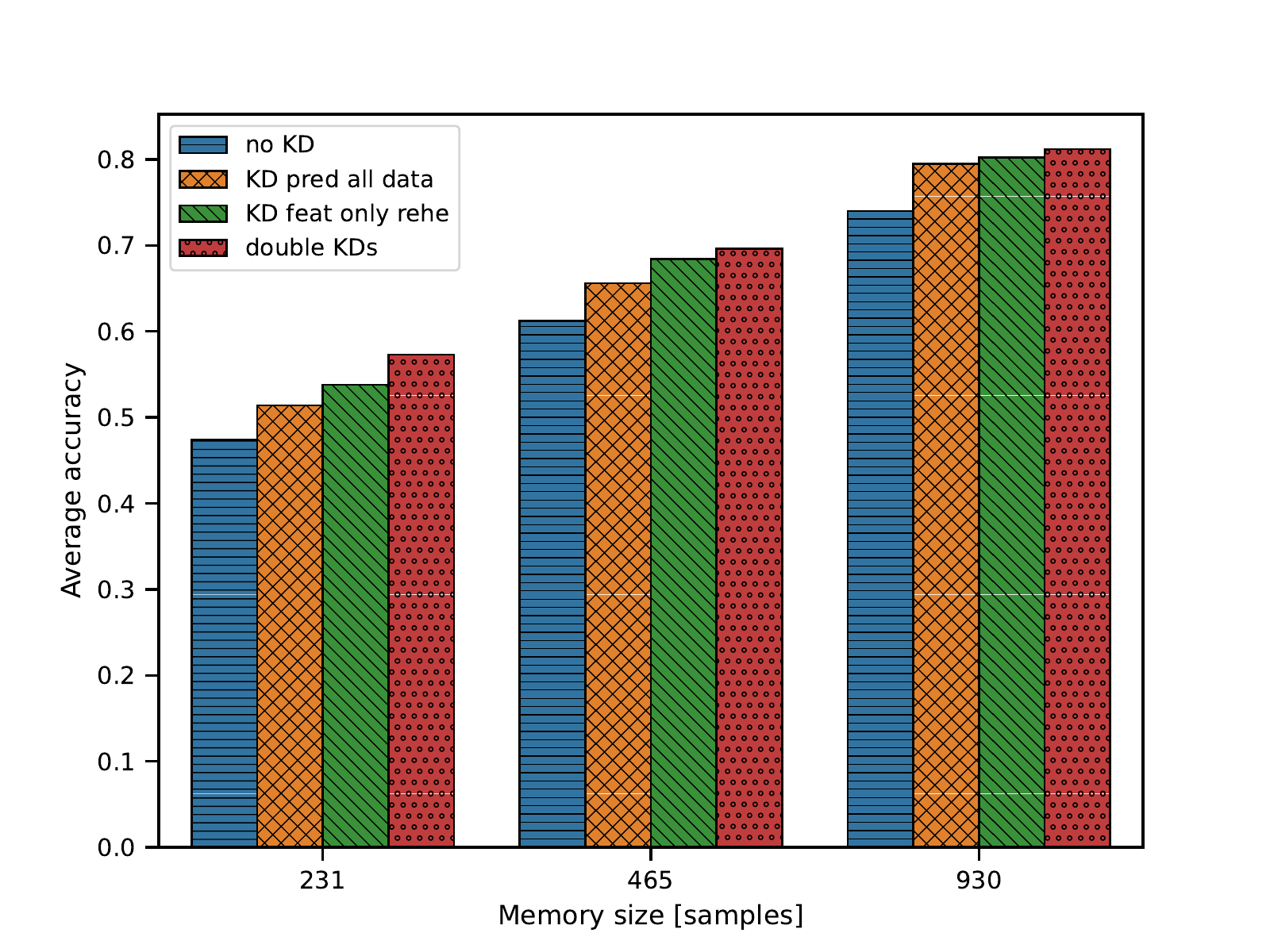}
\caption{Average accuracy for different values of the memory size for the iCaRL strategy.}
\label{fig:bar_plot}
\end{figure}
Fig. \ref{fig:avg_acc_trend} depicts the trend of 4 different configurations for the iCaRL strategy. The concurrent use of both KDs (red curve) leads to the best overall performance, even though the last task accuracy (last acc) is pretty similar to the methods employing single KDs.

Finally, Figure~\ref{fig:bar_plot} shows the average accuracies achieved by different KDs approaches when using smaller rehearsal memory sizes. We can note the consistency of the  performance trends as the memory size changes. In particular, the relative gain in accuracy provided by the KDs is larger when a rehearsal memory with a smaller capacity is used. For instance, the relative gain is around 7.2 points when the memory size is $930$, whereas for memory size = 231 the gain is 9.9. This clearly shows the effectiveness of our approach also for limited-budget SLU systems.

\vspace{-0.3cm}
\section{Conclusions}
\label{sec:conclusion}
 This paper describes an approach for class-incremental continual learning in a SLU domain. We show that the KD on the rehearsal data is effective if applied to the encoded features.
 Furthermore, the feature-level MSE loss, when added to the usual predictions-level KD loss, brings additional performance improvements. The efficacy of the approach is particularly evident when the rehearsal memory size is small, making it suitable for low-resources devices. One limitation is the dataset that, although large in terms of size, lacks lexical richness and variety. Thus, future work will extend the proposed approach to a more recent and complex end-to-end SLU dataset, the Spoken Language Understanding Resource Package (SLURP) \cite{bastianelli2020slurp}, which also features the prediction of multiple entities inside a spoken sentence (e.g., slot filling).

\bibliographystyle{IEEEtran}
\bibliography{mybib}

\end{document}